\documentclass[preprint,showpacs,preprintnumbers,amsmath,amssymb,nofootinbib,showkeys,aps]{revtex4}
\usepackage{graphicx}
\usepackage{dcolumn}
\usepackage{bm}

\usepackage{textgreek}
\usepackage{latexsym}
\usepackage{epsfig}
\usepackage{color}
\usepackage{hyperref}
\usepackage{mathrsfs}
\usepackage{caption}
\usepackage{subfig}
\usepackage{bm}
\usepackage{bm}

\newcommand{\be}{\begin{equation}}
\newcommand{\ee}{\end{equation}}
\newcommand{\beqq}{\setlength\arraycolsep{2pt}\begin{eqnarray}}
\newcommand{\eeqq}{\vspace{0cm} \end{eqnarray}}
\newcommand{\bea}{\begin{eqnarray}}
\newcommand{\eea}{\end{eqnarray}}

\begin{document}

\title{Bayesian correction of $H(z)$ cosmic chronometers data with systematic errors}

\author{N. R. Kvint$^1$}\email{n.kvint@unesp.br}
\author{J. F. Jesus$^{2,1}$}\email{jf.jesus@unesp.br}
\author{S. H. Pereira$^{1}$}\email{shpereira@unesp.br}

\affiliation{$^1$Universidade Estadual Paulista (UNESP), Faculdade de Engenharia e Ci\^encias de Guaratinguet\'a, Departamento de F\'isica - Av. Dr. Ariberto Pereira da Cunha 333, 12516-410, Guaratinguet\'a, SP, Brazil\\
$^2$Universidade Estadual Paulista (UNESP), Campus Experimental de Itapeva - R. Geraldo Alckmin, 519, 18409-010, Itapeva, SP, Brazil
}



\begin{abstract}
We show that the 32 $H(z)$ data from cosmic chronometers have overestimated uncertainties and make use of a Bayesian method to correct and reduce it. We then use the corrected data to constrain flat $\Lambda$CDM and O$\Lambda$CDM parameters. For the flat $\Lambda$CDM model, we got as result $H_{0} = 67.1\pm 4.0$ km s$^{-1}$ Mpc$^{-1}$ and $\Omega _{m} = 0.333 ^{+0.041}_{-0.057}$. While for the O$\Lambda$CDM model, we found $H_{0} = 67.2\pm 4.8$ km s$^{-1}$ Mpc$^{-1}$, $\Omega _{m} = 0.36\pm 0.16$ and $\Omega _{\Lambda} =0.71 ^{+0.36}_{-0.28}$. These results goes from $22\%$ up to $28\%$ uncertainty reduction when compared to the constraints of the both uncorrected models.
\end{abstract}

%
\maketitle


\section{Introduction}

The accelerated expansion of the Universe has been a well-known phenomenon since the first significant evidence emerged from the observation of type Ia supernovae \cite{Riess_1998, Perlmutter_1999}, later corroborated by data from Baryon Acoustic Oscillations (BAO) measurements and Cosmic Microwave Background (CMB) anisotropies. Although these observations were made in 1998, the underlying cause of the accelerated cosmic expansion remains one of the greatest mysteries in modern physics. Following this groundbreaking discovery, a host of potential candidates were proposed to explain the phenomenon. Among them, the cosmological constant $\Lambda$ has garnered significant attention, as it represents the simplest model that aligns well with most current data and observations.

Together with cold dark matter, the cosmological constant forms the foundation of the current standard model of cosmology, known as the $\Lambda$CDM model. Despite its remarkable success in explaining observations, the $\Lambda$CDM model still faces several significant challenges, such as the cosmological constant problem and, more recently, the $H_{0}$ tension \cite{universe9020094, DiValentino_2021}. Consequently, further investigations are essential on the nature and composition of the term $\Lambda$.

To measure the expansion of the Universe directly, three primary methods are currently available: Baryon Acoustic Oscillation (BAO) techniques \cite{10.1111/j.1365-2966.2009.15405.x, Blake_2012, Busca_2013, Anderson_2014, Font_Ribera_2014, Delubac_2015}, the correlation function of luminous red galaxies \cite{Chuang_2013, Oka_2014} and the differential dating of cosmic chronometers \cite{Simon_2005, DanielStern_2010, Moresco_2015, Moresco_2016}. 

For the present analysis, cosmic chronometer measurements of $H(z)$ were chosen. This choice is justified by their model independence and, more importantly, their relatively large uncertainties. An overestimation in the uncertainties in the $H(z)$ data was already discussed in \cite{Jesus_2018}. A method to correct these uncertainties based on \cite{hogg2010data} uses the $\Lambda$CDM as a generative model and a goodness-of-fit criterion, and the result was a reduction of up to 30\% in the uncertainties when compared to the uncorrected analysis \cite{Jesus_2018}. In the present work, differently from \cite{Jesus_2018}, we consider not only the uncertainties of each point, but also their covariances. Thus, we arrive at a new generative model by introducing a correction factor into the covariance matrix. Based on this, the overestimation in the $H(z)$ data is re-evaluated.

This article is organized as follows. In Section \ref{dynamics}, we discuss the basic mechanisms of the $\Lambda$CDM model. Section \ref{Data analysis} presents the 32 $H(z)$ data points used in this study, and in Subsection \ref{Goodness of fit}, we analyze and discuss the goodness of fit. In Section \ref{Method}, we describe how a correction factor can be introduced into the covariance matrix. Subsequently, in Sections \ref{OLCDM} and \ref{FLCDM}, this method is applied to the $\Lambda$CDM model with spatial curvature and to the flat $\Lambda$CDM model, respectively. Section \ref{Comparison} provides a comparison of our results with other $H(z)$ data analyzes and existing estimates of $H_{0}$. Finally, in Section \ref{conclusion}, we summarize our findings and discussions.

\section{\label{dynamics} Dynamics of the ΛCDM model}
In order to analyze if there is an overestimation of the uncertainties and to correct them, we have to assume a fiducial cosmological model. We choose the cosmic concordance $\Lambda$CDM model as a fiducial model, given that it is in agreement with many observational data.

To understand the cosmic dynamics of the standard cosmological model, we begin assuming a homogeneous and isotropic FLRW line element given by:

\begin{equation}
    ds^{2} = -dt^{2} + a^{2}(t)\left[\frac{dr^{2}}{1-kr^{2}} + r^{2}\left(d\theta^{2} + \sin^2\theta d\phi^{2}\right)\right]\,,
    \label{eq:flrw}
\end{equation}
where ($r$, $\theta$, $\phi$) are comoving coordinates, the curvature parameter $k$ can assume the values $-1$, $0$ or $1$, $a(t)$ is the scale factor and $c = 1$. With this metric as the background, Einstein field equations become:

\begin{equation}
    \frac{\dot{a}^{2}}{a^{2}} = \frac{8\pi G}{3} \rho - \frac{k}{a^{2}}
    \label{eq:friedmann}
\end{equation}
\begin{equation}
    \frac{\ddot{a}}{a} = -\frac{4\pi G}{3}(\rho + 3p)
\end{equation}
where $\rho$ is the total energy density and $p$ is the total pressure of the cosmic fluid. By taking the time derivative of the first equation and using the second one we obtain:
\begin{equation}
    \dot{\rho} + 3\frac{\dot{a}}{a}(\rho + p) =0\,,\label{rhot}
\end{equation}
which describes the evolution of the energy density given an equation of state $p=w\rho$.

As we know, the cosmological fluid can consist of different components. For radiation, we have $w = 1/3$, and thus Eq. (\ref{rhot}) yields $\rho_{r} \propto a^{-4}$. For ordinary baryonic matter as well as dark matter, we have $w\approx 0$, so that $p_{m} \approx 0$ and its density evolves with $\rho_{m} \propto a^{-3}$. For a constant $\Lambda$, we have $w=-1$, so that $\rho _{\Lambda} = \rho _{\Lambda 0}$, a constant. By introducing the Hubble parameter $H(t)=\frac{\dot{a}}{a}$ and the redshift $z$ through $a=\frac{1}{1+z}$, the first Friedmann equantion (\ref{eq:friedmann}) turns:
\begin{equation}
    H^{2} = \frac{8\pi G}{3} [\rho _{r0}(1+z)^{4} + \rho _{m0}(1+z)^{3} +\rho _{\Lambda 0}]- {k(1+z)^{2}}\,,\label{H2}
\end{equation}
where $\rho_{r0}$ and $\rho_{m0}$ represent the present-day values of the radiation and total matter densities, respectively.

It is well known that radiation is negligible in the present epoch; therefore, it will be disregarded. Furthermore, by defining the critical energy density $\rho _{c0} = \frac{3H_{0}^{2}}{8\pi G}$ and the density parameter $\Omega _{i} = \frac{\rho _{i0}}{\rho_{c0}}$ for each component, we have:

\begin{equation}
    \left( \frac{H}{H_{0}} \right) ^{2} = \Omega _{m}(1+z)^{3} + \Omega _{k} (1+z)^{2} + \Omega _{\Lambda}
    \label{eq:friedmannW}
\end{equation}
where $\Omega_k\equiv-\frac{k}{a_0^2H_0^2}$ and we are assuming $a_0=1$. From \eqref{eq:friedmannW} we can finally obtain an expression for $H(z)$:
\begin{equation}
    H(z) = H_{0} \left[ \Omega _{m}(1+z)^{3} + (1-\Omega _{m}-\Omega _{\Lambda}) (1+z)^{2} + \Omega _{\Lambda} \right]^{1/2}\,,
    \label{eq:Hz}
\end{equation}
where we have used the normalization $\Omega _{m} + \Omega _{k} + \Omega _{\Lambda} = 1$ from \eqref{eq:friedmannW} for $z=0$. Thus we are left to a three parameters model: $H_0$, $\Omega_m$ and $\Omega_\Lambda$. This model will be named O$\Lambda$CDM since it is allowed some curvature. 

In contrast, for the flat $\Lambda$CDM model, $\Omega _{k} = 0$, we have:
\begin{equation}
    H(z) = H_{0} \left[ \Omega _{m}(1+z)^{3} + 1 - \Omega _{m} \right]^{1/2}
    \label{eq:flatHz}\,,
\end{equation}
with just two free parameters: $H_0$ and $\Omega_m$. 

In the following, we will analyze both models.

\section{\label{Data analysis} Data Analysis}

The data used in this work consist solely of $H(z)$ measurements obtained through cosmic chronometers, as they are independent of cosmological models and rely exclusively on astrophysical assumptions. The dataset contains 32 $H(z)$ measurements, which are presented in \cite{Moresco_2022} and summarized in Tab. \ref{tabHz}. This is an up-to-date compilation where, for the first time, Moresco \textit{et al.} estimated systematic errors for the $H(z)$ data, described by a covariance matrix.\makeatletter\def\Hy@Warning#1{}\makeatother\footnote{The method for deriving the covariance matrix can be found at \url{https://gitlab.com/mmoresco/CCcovariance}.}

\begin{table}[ht]
\centering
\begin{tabular}{ccc}
\hline
$z$    & ~~~~$ H(z) $~~~~ & $\sigma _{H}$ \\
\colrule
0.070  ~~~~& 69     &~~~~ 19.6    ~~~~      \\
0.090  ~~~~& 69     &~~~~ 12      ~~~~      \\
0.120  ~~~~& 68.6   &~~~~ 26.2    ~~~~      \\
0.170  ~~~~& 83     &~~~~ 8       ~~~~      \\
0.1791 ~~~~& 75     &~~~~ 4       ~~~~      \\
0.1993 ~~~~& 75     &~~~~ 5       ~~~~      \\
0.200  ~~~~& 72.9   &~~~~ 29.6    ~~~~      \\
0.270  ~~~~& 77     &~~~~ 14      ~~~~      \\
0.280  ~~~~& 88.8   &~~~~ 36.6    ~~~~      \\
0.3519 ~~~~& 83     &~~~~ 14      ~~~~      \\
0.3802 ~~~~& 83     &~~~~ 13.5    ~~~~      \\
0.400  ~~~~& 95     &~~~~ 17      ~~~~      \\
0.4004 ~~~~& 77     &~~~~ 10.2    ~~~~      \\
0.4247 ~~~~& 87.1   &~~~~ 11.2     ~~~~     \\
0.445  ~~~~& 92.8   &~~~~ 12.9    ~~~~      \\
0.470  ~~~~& 89     &~~~~ 50      ~~~~      \\
\hline
\end{tabular}
\quad
\begin{tabular}{ccc}
\hline
$z$    & ~~~~$ H(z) $~~~~ & $\sigma _{H}$ \\
\colrule
0.4783  ~~~~& 80.9   &~~~~ 9      ~~~~       \\
0.480   ~~~~& 97     &~~~~ 62     ~~~~       \\
0.5929  ~~~~& 104    &~~~~ 13     ~~~~       \\
0.6797 ~~~~& 92     &~~~~ 8      ~~~~        \\
0.750   ~~~~& 98.8   &~~~~ 33.6   ~~~~       \\
0.7812  ~~~~& 105    &~~~~ 12     ~~~~       \\
0.8754  ~~~~& 125    &~~~~ 17     ~~~~       \\
0.880   ~~~~& 90     &~~~~ 40     ~~~~       \\
0.900   ~~~~& 117    &~~~~ 23     ~~~~       \\
1.037   ~~~~& 154    &~~~~ 20     ~~~~       \\
1.300   ~~~~& 168    &~~~~ 17     ~~~~       \\
1.363   ~~~~& 160    &~~~~ 33.6   ~~~~       \\
1.430   ~~~~& 177    &~~~~18      ~~~~       \\
1.530   ~~~~& 140    &~~~~ 14     ~~~~       \\
1.750   ~~~~& 202    &~~~~ 40     ~~~~       \\
1.965   ~~~~& 186.5  &~~~~ 50.4   ~~~~       \\
\hline
\end{tabular}
\caption{The 32 $H(z)$ versus $z$ data used in this work.}
\label{tabHz}
\end{table}

\begin{figure}[ht]
\centering
    \includegraphics[width=0.7\textwidth]{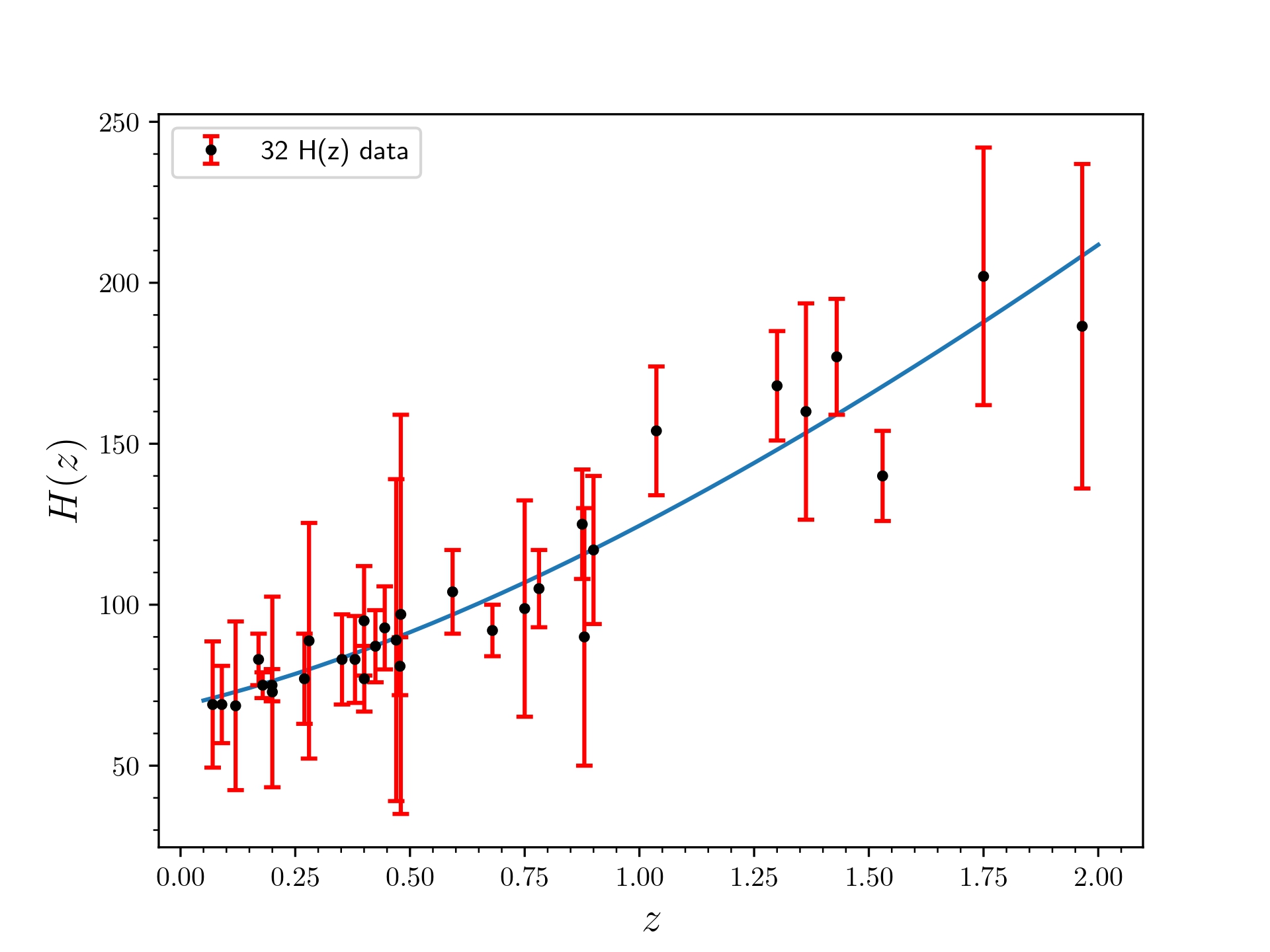}
    \caption{Best fit to the 32 $H(z)$ data for the flat $\Lambda$CDM model.}
    \label{Hzdata}
\end{figure}

Since these data is given in terms of a covariance matrix $\Sigma$, by applying a $\chi^{2}$-statistic to them, we obtain the following function of free parameters:
\begin{equation}
    \chi^{2} = \sum_{i,j}^{32}[H_{obs,i} - H_{mod}(z_i,\vec{\theta})]\Sigma^{-1}_{ij}[H_{obs,j} - H_{mod}(z_j,\vec{\theta})]
    \label{eq:uncorrX2}
\end{equation}
where $H_{obs,i}$ represents the $H(z)$ data, $H_{mod}(z_i)$ is the $H(z)$ value predicted by the model evaluated at redshift $z_i$, and $\vec{\theta} \equiv (H_0, \Omega_m, \Omega_\Lambda)$ is the vector of cosmological parameters. The 32 $H(z)$ data points, along with the best-fit curve for the flat $\Lambda$CDM model, are shown in Fig. \ref{Hzdata}.

\subsection{\label{Goodness of fit} Goodness of fit}
One possible way to evaluate the goodness of fit is by examining the $\chi^2_\nu$ probability density function (PDF). The reduced $\chi^2$, $\chi^2_\nu$, is defined as $\chi^2_\nu = \chi^2_{\text{min}} / \nu$, where $\nu = n - p$ represents the number of degrees of freedom, with $n$ being the number of data points and $p$ the number of free parameters. The $\chi^2_\nu$ distribution is given by \cite{bevington2003data, vuolo}:

\begin{equation}
    h_{\nu} (\chi ^{2} _{\nu}) = \frac{\nu ^{\frac{\nu}{2}}(\chi _{\nu} ^{2})^{\frac{1}{2}(\nu - 2)}e^{-\frac{\nu}{2}\chi ^{2} _{\nu}}}{2^{\frac{\nu}{2}}\Gamma \left( \frac{\nu}{2} \right)}
    \label{eq:hv}
\end{equation}
where $\Gamma$ is the complete gamma function. It is known that the expected value of $\chi^2_\nu$ is 1 \cite{bevington2003data, vuolo}; therefore, values significantly deviating from this are considered unlikely. High values may suggest a poor fit to the data or an underestimation of uncertainties, while low values generally indicate an overestimation of uncertainties.

We can also define the cumulative distribution function (CDF), which give us the probability of obtaining a $\chi _{\nu}^{2}$ value as low as $Q$, by:
\begin{equation}
    P(\chi _{\nu} ^{2} < Q) \equiv \int _{0} ^{Q} h_{\nu}(Q')dQ'\,.
    \label{eq:cdf}
\end{equation}

We are now able to access the goodness of fit for the O$\Lambda$CDM and flat $\Lambda$CDM models. For the $\chi _{\nu} ^{2}$ value, we have:
\begin{equation}
    \chi _{\nu} ^{2} =
    \begin{cases}
        \frac{14.475}{29} = 0.49915\text{, for O$\Lambda$CDM} \\
        \frac{14.533}{30} = 0.48442\text{, for Flat $\Lambda$CDM}
    \end{cases}
\end{equation}

\begin{figure}[ht]
  \centering
  \subfloat[O$\Lambda$CDM.]{\includegraphics[width=0.5\textwidth]{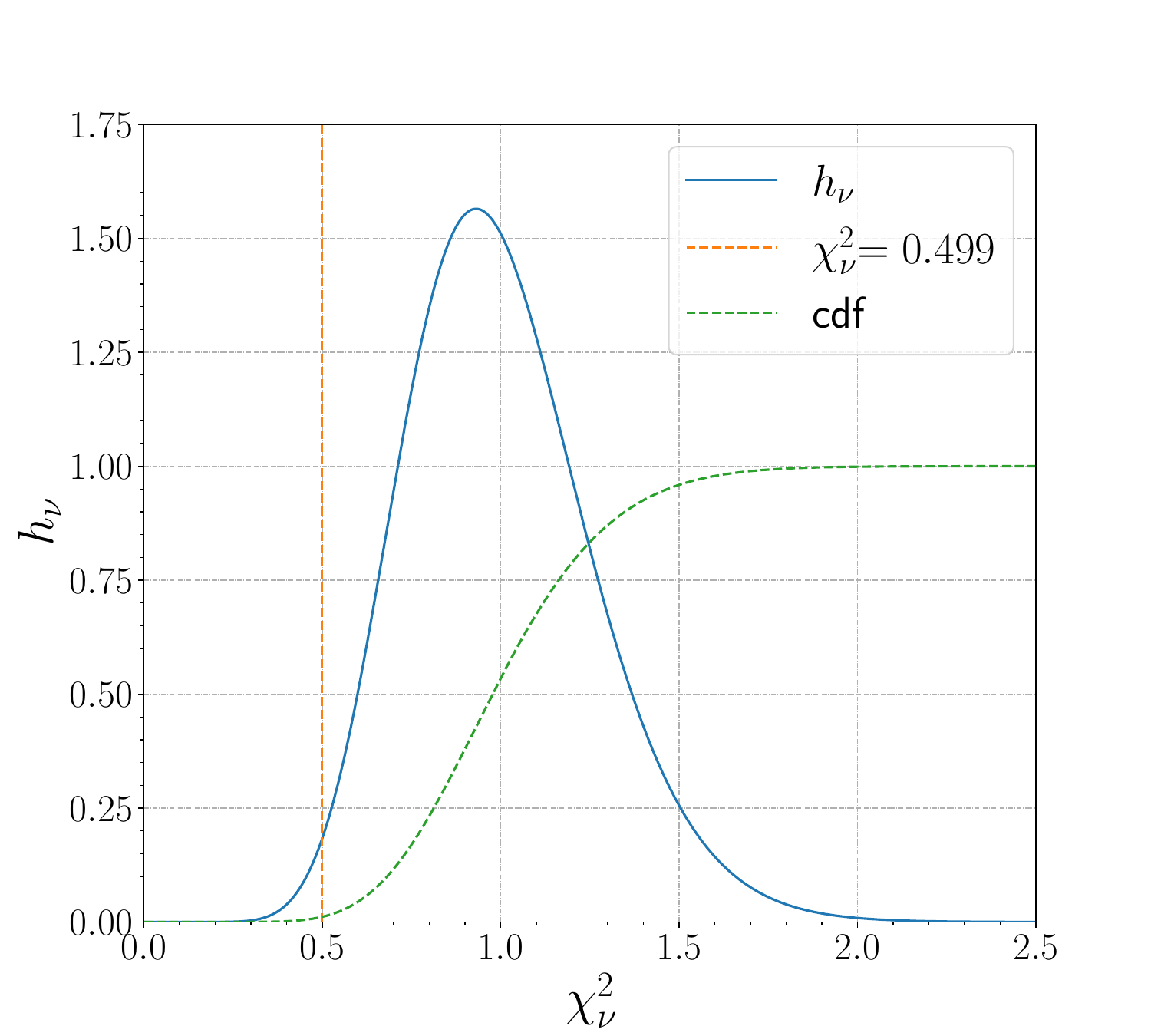}\label{fig:OLcdm}}
  \hfill
  \subfloat[Flat $\Lambda$CDM.]{\includegraphics[width=0.5\textwidth]{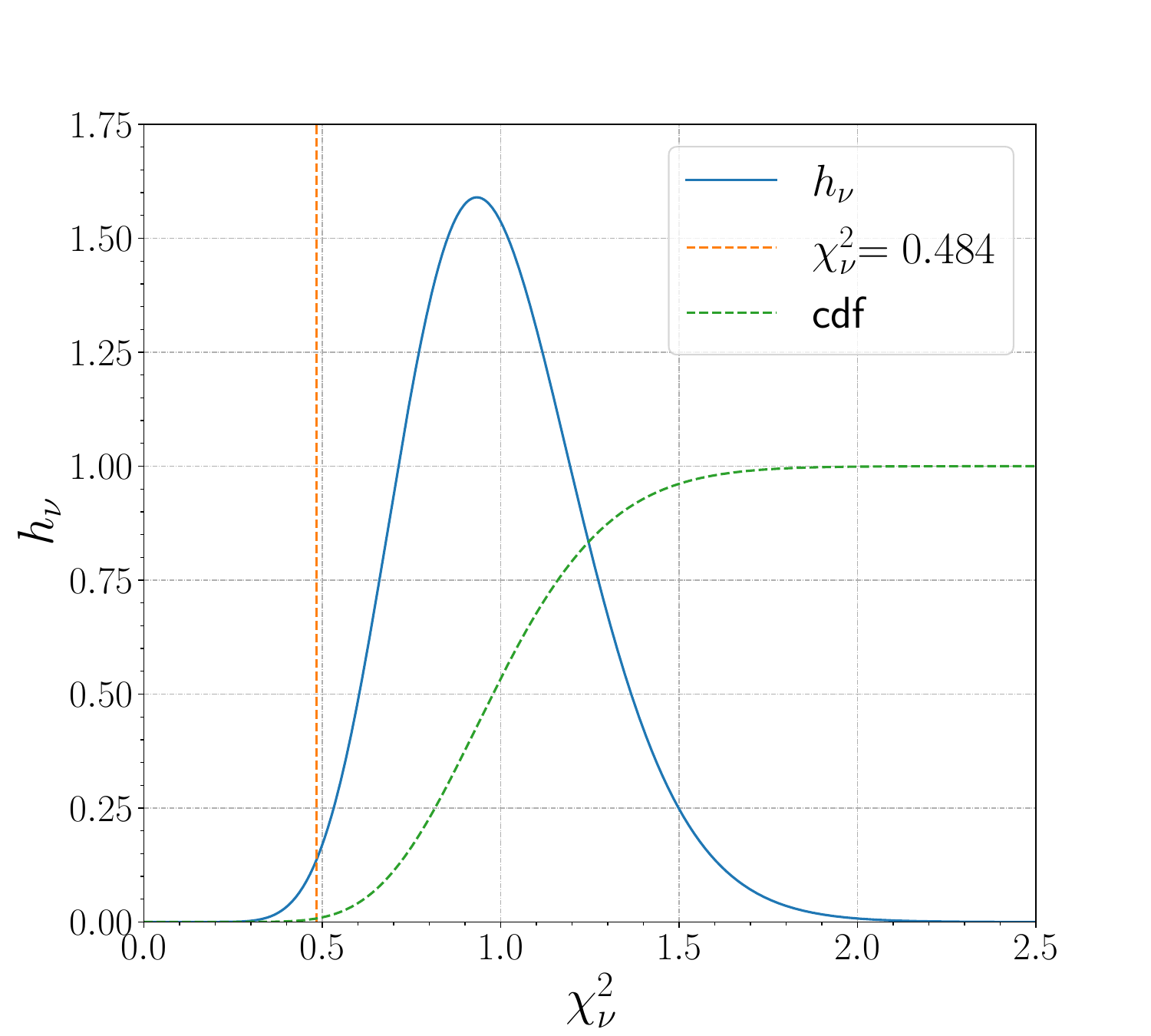}}\label{fig:FLcdm}
  \captionof{figure}{$h_{\nu}(\chi _{\nu} ^{2})$ and its corresponding cdf. For (a), $\nu = 29$, and for (b), $\nu = 30$.}
  \label{fig:hv}
\end{figure}

Figure \ref{fig:hv} shows that these two obtained values appear to be quite unlikely. By calculating each CDF using \eqref{eq:cdf}, we find that $P(\chi^2_\nu < 0.49915) = 1.13\%$ and $P(\chi^2_\nu < 0.48442) = 0.79\%$. Thus, it is evident that our $\chi^2_\nu$ values are both highly unlikely. This result suggests that the low $\chi^2$ values point to an overestimation of the $H(z)$ uncertainties.

\section{\label{Method} How to correct the $H(z)$ uncertainties?}

Now that we have evidence suggesting overestimated uncertainties, we can begin to consider how these might be corrected. Typically, we work with a dataset, in this case ($z_{i}, H_{i}, \sigma_{H_i}$), and aim to construct a likelihood function of the form $\mathcal{L} \propto e^{-\chi^2 / 2}$. More specifically, we seek to maximize this likelihood, which is equivalent to minimizing $\chi^2$. To achieve this, we need to determine the parameter values that provide the best fit of the model to the data.

One approach to accomplish this is by sampling the posterior distribution, which is proportional to the likelihood under the assumption of flat priors for the parameters. By sampling the posterior, we can not only identify the best-fit parameters but also estimate their mean, median, and uncertainties. However, in this work, our primary focus is to introduce a correction to the $H(z)$ uncertainties, which are encapsulated in the covariance matrix.

Similarly to \cite{Jesus_2018}, and as mentioned in \cite{hogg2010data}, the most effective approach is to construct a generative model for the data. This can be understood as a statistical framework designed to reproduce the observed data. With these concepts in mind, we can begin applying our corrections. For a $k$-dimensional multivariate normal distribution, the likelihood is given by
\begin{equation}
    \mathcal{L} = (2\pi)^{-k/2} \text{det}(\Sigma) ^{-1/2} e ^{-\chi ^{2}/2}\,.
    \label{eq:oldL}
\end{equation}

The most rigorous way to correct the uncertainties would be to apply a correction factor to each individual data point. However, since we aim to avoid introducing more free parameters than data points, we will apply the same correction factor $f$ to all uncertainties, as in \cite{Jesus_2018}. Unlike \cite{Jesus_2018}, however, we not only have the uncertainties for each data point, but also their covariances. In this case, the correction factor must be applied in such a way that it reproduces the same correction as in \cite{Jesus_2018} in the particular case where the data are uncorrelated. To achieve this, we must introduce a $f^{2}$ factor for each element of the covariance matrix. This ensures that the correlation between data points remains unchanged while applying the correction. Thus, we can write
\begin{equation}
    \Sigma_{corr} = f^{2} \Sigma\,,
    \label{eq:corrCov}
\end{equation}
where $\Sigma_{corr}$ is the corrected covariance matrix.

Obtaining $\Sigma_{corr} ^{-1}$ is immediate and can be easily checked to be true:
\begin{equation}
    \Sigma_{corr} ^{-1} = f^{-2}\Sigma ^{-1}\,.
    \label{eq:corrInvCov}
\end{equation}

Now we must to check how the equations \eqref{eq:oldL} and \eqref{eq:uncorrX2} will be modified. Starting with \eqref{eq:uncorrX2}, we have that the corrected $\chi ^{2}$ is:
\begin{equation}
    \chi ^{2} _{corr} = \sum_{i,j}^{32}[H_{obs,i} - H_{mod}(z_i)]\Sigma^{-1}_{corr, ij}[H_{obs,j} - H_{mod}(z_j)] = f^{-2} \chi ^{2}
\end{equation}
We conclude that the corrected $\chi ^{2}$ is proportional to the original one by a factor of $f^{-2}$.

Going now to the equation \eqref{eq:oldL}:
\begin{equation}
    \mathcal{L}_{corr} = (2\pi)^{-k/2} \text{det}(\Sigma _{corr}) ^{-1/2} e ^{-\chi _{corr} ^{2}/2} = (2\pi)^{-k/2} \text{det}(f^{2}\Sigma) ^{-1/2} e^{-f^{-2} \chi ^{2}/2}\,,
    \label{eq:corrL}
\end{equation}
since the covariance matrix has dimension $k$, we can say that
\begin{equation}
    \text{det} (f^{2} \Sigma) = f^{2k} \text{det}(\Sigma)\,,
\end{equation}
and by taking the logarithm of \eqref{eq:corrL}, we finally get:
\begin{equation}
    \text{ln}(\mathcal{L}_{corr}) = A -k\text{ln}(f) - f^{-2}\frac{\chi ^{2}}{2}
\end{equation}
where $A = -\frac{1}{2}\ln\left[ (2\pi)^k\det(\Sigma) \right]$. As the likelihood is given by $\mathcal{L} = Ne^{-\chi ^{2}/2} $, and $N$ is a normalization constant, it can absorb the term given by $A$.

Now that we have an expression for the likelihood, we simply need to maximize it to obtain the best-fit cosmological parameters and the optimal correction factor $f$. These two results, taken together, will provide us with the best model to describe the data.

To find the constraints on the parameters $H_{0}$, $\Omega_{m}$, $\Omega_{\Lambda}$, and $f$, we sampled the likelihood $\mathcal{L}$ using a Monte Carlo Markov Chain (MCMC) analysis, specifically employing the Affine Invariant MCMC Ensemble Sampler \cite{goodman2010} (see \cite{Jesus_2018} for a brief justification of the use of this method). A {\sffamily Python} implementation was developed using the {\sffamily emcee} software \cite{Foreman_Mackey_2013}.

\section{\label{Correction}$H(z)$ UNCERTAINTIES CORRECTION: Results}

\renewcommand{\arraystretch}{1.25}

\subsection{\label{FLCDM}Flat ΛCDM}

\begin{figure}[ht]
    \centering
    \includegraphics[width=0.75\textwidth]{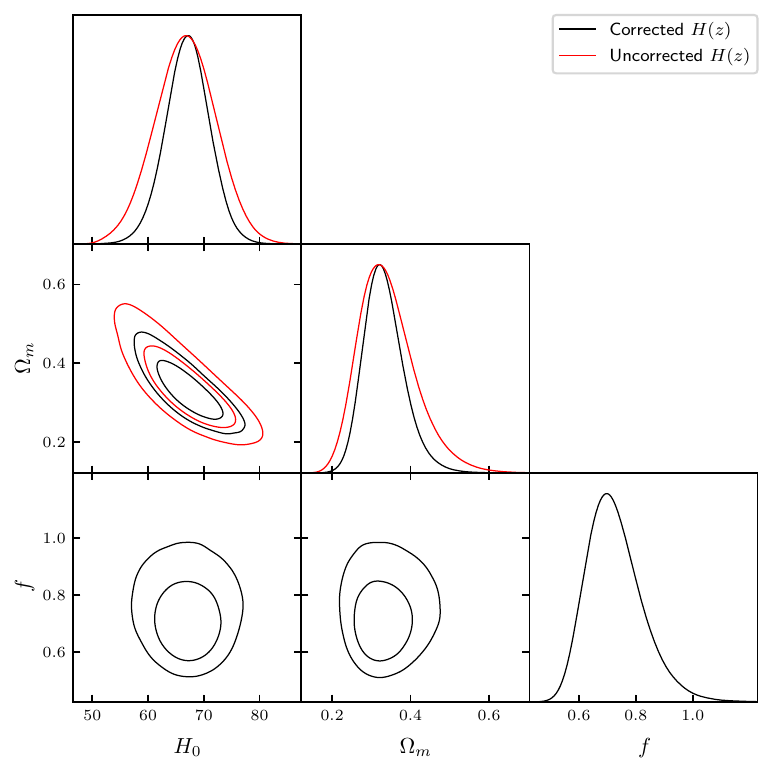}
    \caption{Results of the analysis of the flat $\Lambda$CDM model, with $H_{0}$ in km/s/Mpc. The diagonal contains the marginalized constraints for each parameter. Below the diagonal are the marginalized contours (68.3\% and 95.4\% confidence levels) for each pair of parameters.}
    \label{fig:contoursFlat}
\end{figure}

As shown in Figure \ref{fig:contoursFlat}, the introduction of the correction parameter $f$ modifies the range of values that the cosmological parameters can take. The corrected contours (black) are more constrained than the uncorrected ones (red). Another observation is that the introduction of $f$ has minimal impact on the correlation between the parameters, resulting in only a small change in their mean values.

Table \ref{tab:FLCDM} shows the results obtained for the mean values of the parameters: all uncertainties were reduced after the correction was applied. Specifically, $\sigma _{H0}$ decreased from 5.4 to 4.0, and $\sigma _{\Omega {m}}$ changed from 0.054 and $-0.084$ to 0.041 and $-0.057$. The correction factor has a mean value of $f = 0.727^{+0.073}_{-0.11}$.

\begin{center}
    \begin{tabular} { c c c}
    & Flat $\Lambda$CDM &\\
    \hline
     Parameter & ~~~~~Uncorrected~~~~~ & Corrected\\
    \hline
    $\bm{H_0}$ & $66.7\pm 5.4 \pm 10.0$ & $67.1\pm 4.0 \pm8.0$\\
    
    $\bm{\Omega_m}$ & $0.342^{+0.054+0.150}_{-0.084-0.130}$ & $0.333^{+0.041+0.100}_{-0.057-0.097}$\\
    
    $\bm{f}$ & ----- & $0.728^{+0.073+0.200}_{-0.110-0.180}$\\
    \hline
    \end{tabular}
    \captionof{table}{Mean values of the parameters for flat $\Lambda$CDM model from $H(z)$ data without and with correction factor $f$ on the uncertainties. The uncertainties correspond to 68\% and 95\%  confidence level.}
    \label{tab:FLCDM}
\end{center}

\subsection{\label{OLCDM}OΛCDM}

\begin{figure}[ht]
    \centering
    \includegraphics[width=0.9\textwidth]{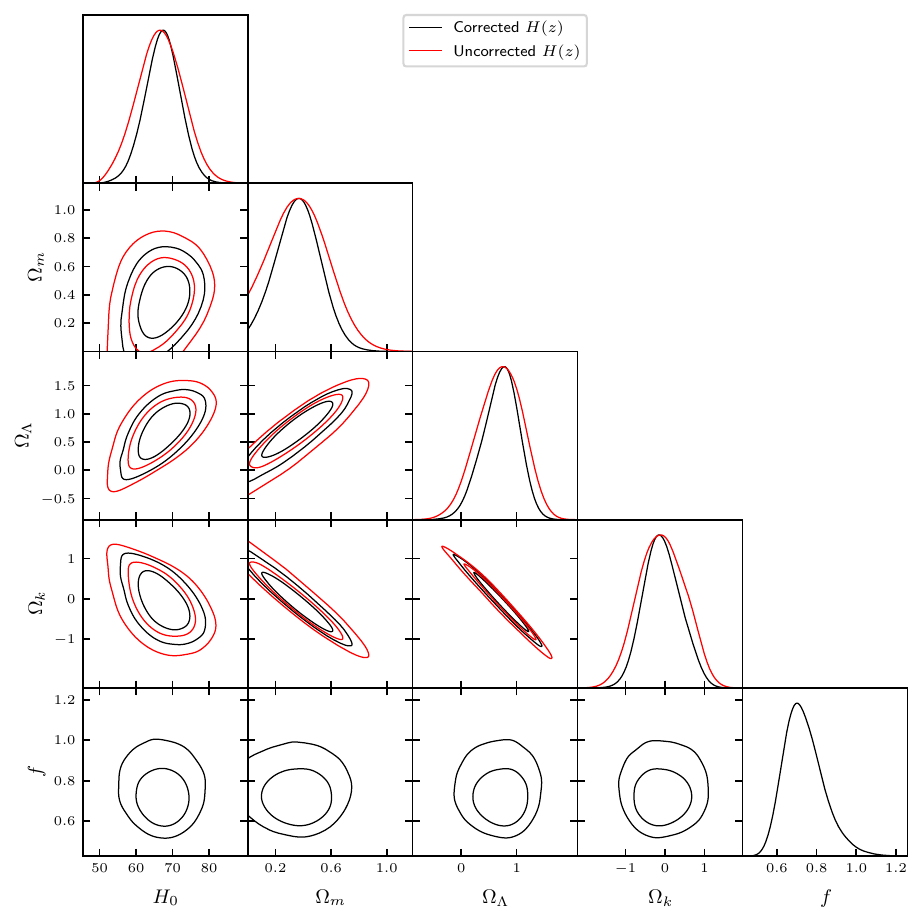}
    \caption{Results of the analysis of the O$\Lambda$CDM model, with $H_{0}$ in km/s/Mpc. The diagonal contains the marginalized constraints for each parameter. Below the diagonal are the marginalized contours (68.3\% and 95.4\% confidence levels) for each pair of parameters.}
    \label{fig:contoursO}
\end{figure}

Looking at the O$\Lambda$CDM model, we observe similar results. Figure \ref{fig:contoursO} shows that, once again, the corrected contours (black) are more restricted than the uncorrected ones (red). The correction factor still does not show any significant correlation with the parameters. It is important to note that the $\Omega_k$ parameter is a derived one, obtained through normalization, as shown in expression \ref{eq:friedmannW}.

Table \ref{tab:OLCDM} shows the results obtained for the mean values of the parameters. Again, all uncertainties were reduced after the correction was applied: $\sigma_{H0}$ decreased from 6.2 to 4.8, $\sigma_{\Omega_{m}}$ changed from 0.19 and $-0.23$ to $\pm 0.16$, $\sigma_{\Omega_{\Lambda}}$ changed from 0.44 and $-0.38$ to 0.36 and $-0.28$, and $\sigma_{\Omega_{k}}$ decreased from 0.6 to 0.43 and $-0.51$. The correction factor has a mean value of $f = 0.74^{+0.06}_{-0.12}$.

\begin{center}
    \begin{tabular} { c c c}
    & O$\Lambda$CDM &\\
    \hline
     Parameter & ~~~~~Uncorrected~~~~~ & Corrected\\
    \hline
    $\bm{H_0}$ & $66.6\pm 6.2\pm 10.0$ & $67.2 ^{+4.8+9.0}_{-4.8-10.0}$\\
    
    $\bm{\Omega_m}$ & $0.38^{+0.19+0.34}_{-0.23-0.34}$ & $0.36^{+0.16+0.29}_{-0.16-0.32}$\\
    
    $\bm{\Omega_\Lambda}$ & $0.70^{+0.44+0.74}_{-0.38-0.81}$ & $0.71^{+0.36+0.58}_{-0.28-0.67}$\\
    
    $\Omega_k$ & $-0.1\pm 0.6\pm1.1$ & $-0.07^{+0.43+0.94}_{-0.51-0.87}$\\
    
    $\bm{f}$ & ----- & $0.74^{+0.06+0.20}_{-0.12-0.18}$\\
    \hline
    \end{tabular}
    \captionof{table}{Mean values of the parameters for O$\Lambda$CDM model from $H(z)$ data without and with correction factor $f$ on the uncertainties. Parameters in bold text are primitive ones, while the other are derived. The uncertainties correspond to 68\% and 95\% confidence level.}
    \label{tab:OLCDM}
\end{center}

\subsection{\label{Comparison}Comparison with other analyses and results}

A similar analysis was conducted by Jesus \textit{et al.} \cite{Jesus_2018}, where they corrected uncertainties from an $H(z)$ data compilation, including both cosmic chronometers and BAO. In contrast, we have chosen to work exclusively with cosmic chronometers, as they are independent of any fiducial cosmological model. Another difference is that, in the present work, systematic errors are now taken into account, as estimated by \cite{Moresco_2022}. Despite these differences, their analysis is similar to ours, and we can use their results to assess how the systematic errors impact the analysis.

For the flat $\Lambda$CDM model, Jesus \textit{et al.} \cite{Jesus_2018} found $H_{0} = 70.3 \pm 1.7$ km s$^{-1}$ Mpc$^{-1}$ for uncorrected $H(z)$ data, and $H_{0} = 70.4 \pm 1.2$ km s$^{-1}$ Mpc$^{-1}$ for the corrected data. These results are in agreement with our findings in Table \ref{tab:FLCDM}. For the O$\Lambda$CDM model, they found $H_{0} = 69.1 \pm 3.5$ km s$^{-1}$ Mpc$^{-1}$ for uncorrected $H(z)$ data, and $H_{0} = 69.5 \pm 2.5$ km s$^{-1}$ Mpc$^{-1}$ for the corrected ones. This is also consistent with our results shown in Table \ref{tab:OLCDM}.

Since there is a tension between different estimations of $H_{0}$ from early and late data, it is valid to assess how compatible our results are with two of the main estimates. Firstly, Breuval \textit{et al.} (2024) \cite{Breuval_2024} (SH0ES) obtained $H_{0} = 73.17 \pm 0.86$ km s$^{-1}$ Mpc$^{-1}$. Our result in Table \ref{tab:FLCDM} is compatible with theirs at the 1.4$\sigma$ confidence level, while our result in Table \ref{tab:OLCDM} is compatible at the 1.2$\sigma$ confidence level with the SH0ES results. The other relevant estimate for the $H_{0}$ tension comes from the Planck Collaboration (2018) \cite{planck2018}. In their study, they found $H_{0} = 67.36 \pm 0.54$ km s$^{-1}$ Mpc$^{-1}$. This result is, more comfortably, compatible with our result from Table \ref{tab:FLCDM} at the 0.06$\sigma$ confidence level.

This result is in agreement with high-redshift CMB data from Planck 2018, in the context of the $\Lambda$CDM model, but it is in tension with the local Cepheid and SNe Ia distance ladder measurements. This is surprising, as these data are located at low redshift and are independent of cosmological models, so one would not expect such a discrepancy. However, our results for the uncorrected $H(z)$ data show larger uncertainties, such that our estimated $H_0$ value is in agreement with both early- and late-time data constraints. This suggests that accounting for systematic errors is indeed necessary.

\section{\label{conclusion}Conclusions}

In this work, we have proposed an alternative approach to address the overestimation of uncertainties in the Hubble parameter $H(z)$ and density parameters. By generalizing the method presented in \cite{Jesus_2018}, where a reduction of up to 30\% in the uncertainties was obtained when compared to the uncorrected case, we introduced a refinement by considering not only the uncertainties of each data point but also their covariances. This new approach incorporated the systematic errors through the full covariance matrix and was applied to 32 $H(z)$ data points, without relying on any $H_0$ priors or data from other measurement techniques.

Our results demonstrated that this method can slightly reduce the uncertainties in cosmological parameters while simultaneously providing a general correction factor for both the flat $\Lambda$CDM and O$\Lambda$CDM models. Specifically, for the flat $\Lambda$CDM model, we achieved a reduction of approximately 26\% in the uncertainty of $H_0$ and 29\% in the uncertainty of $\Omega_m$. For the O$\Lambda$CDM model, the reductions were around 23\% in the uncertainty of $H_0$ and 20\% for $\Omega_m$.

These results show the effectiveness of our proposed method in refining the constraints on cosmological parameters. Furthermore, such a method for re-evaluating uncertainties can contribute significantly to the ongoing discussion surrounding the Hubble tension, offering an additional perspective by improving the accuracy of $H(z)$ data analyses.

\begin{acknowledgments}
This study was financed by the Coordenação de Aperfeiçoamento de Pessoal de Nível Superior - Brasil (CAPES) - Finance Code 001. JFJ acknowledges financial support from Conselho Nacional de Desenvolvimento Científico e Tecnológico (CNPq) (No. 314028/2023-4). SHP acknowledges financial support from Conselho Nacional de Desenvolvimento Científico e Tecnológico (CNPq) (No. 308469/2021).
\end{acknowledgments}

\nocite{*}
\bibliography{refs}
\end{document}